\documentclass[12pt]{article}
\usepackage[utf8]{inputenc}
\usepackage{graphicx}
\usepackage{amsmath, amssymb}
\usepackage{setspace}
\usepackage[margin=1in]{geometry}
\usepackage[numbers,sort&compress]{natbib}
\usepackage[colorlinks=true,linkcolor=blue,citecolor=blue,urlcolor=blue]{hyperref}
\usepackage{bm}
\title{\textbf{Learning to Maximize Quantum Neural Network Expressivity via Effective Rank}}
\author{
Juan Yao\textsuperscript{a,b,c}\thanks{\texttt{juanyao.physics@gmail.com}}
\\
\textsuperscript{a} International Quantum Academy, Shenzhen, 518048, Guangdong, China
\\
\textsuperscript{b} Qosmos Technology (Beijing) Co., Ltd., Beijing, 100176, China
\\
\textsuperscript{c} Intelligent Quantum Inception Co., Ltd., Haidian, Beijing 100083, China
}
\date{}

\begin{document}
\maketitle
\begin{abstract}
{
Quantum neural networks (QNNs) are widely used as trainable models for solving variational problems, where their ability to represent complex functions directly determines performance. However, accurately quantifying this expressivity remains a major challenge, limiting the understanding of the expressive power of QNNs. 
Here, we introduce the effective rank $\kappa$ as a quantitative measure of expressivity. We show that the expressivity of a QNN is determined by the interplay of three factors: the dataset, the measurement operators, and the circuit structure. Unlike conventional metrics based on parameter sampling or entanglement structure, $\kappa$ captures the number of independent parameters that effectively contribute to the model output, providing an operational and data-dependent characterization of expressivity while avoiding costly parameter sampling. 
We demonstrate that $\kappa$ can reach its theoretical maximum $4^n - 1$ for an $n$-qubit system when the circuit architecture, input distribution, and measurement protocol are jointly optimized. Leveraging this insight, we employ $\kappa$ as a design objective within a reinforcement learning framework with a self-attention transformer agent to automatically discover highly expressive circuit architectures. Our results show that the proposed approach efficiently identifies high-performing circuit configurations, outperforming heuristic designs and random search in terms of sample efficiency.
By bridging theoretical characterization with automated design, our approach provides a practical, scalable, and computationally efficient route for constructing expressive quantum circuits, advancing the development of quantum machine learning models.
}
\end{abstract}

\section{Introduction}
Quantum neural networks (QNNs) have emerged as powerful tools in quantum machine learning~\cite{biamonte_quantum_2017, cerezo_challenges_2023, wang_comprehensive_2024, du_quantum_2025, zaman_survey_2025} and variational quantum algorithms~\cite{benedetti_parameterized_2019, cerezo_variational_2021,  de_palma_limitations_2023, qi_variational_2024}, playing a pivotal role in diverse applications,  including  quantum chemistry~\cite{ lourenco_qmlmaterialquantum_2023}, quantum sensing~\cite{degen_quantum_2017} and financial modeling~\cite{orus_quantum_2019}. 
As parameterized quantum circuits (PQC), QNNs serve as ansätze for optimizing cost functions over high-dimensional Hilbert spaces. Their ability to efficiently represent complex functions directly affects their performance in solving variational problems. Enhancing the performance and scalability of QNNs is a key focus of ongoing research.
Various parameterized quantum circuits (PQCs) have been proposed to enhance the expressivity and effectiveness of quantum neural networks (QNNs) in diverse applications. These include convolutional QNNs~\cite{cong_quantum_2019, henderson_quanvolutional_2020, herrmann_realizing_2022}, recurrent QNNs~\cite{bausch_recurrent_2020, li_quantum_2023}, quantum generative adversarial networks (QGANs)~\cite{huang_experimental_2021,PhysRevLett.121.040502,NEURIPS2019_f35fd567}, and randomness-enhanced QNNs\cite{wu_randomness-enhanced_2024}. Each of these architectures leverages unique design principles to improve representation power and tackle complex quantum learning tasks.

Various metrics have been explored to assess and compare the expressivity of different quantum neural network (QNN) architectures. In Ref.~\cite{sim_expressibility_nodate}, a statistical measure, Expr, was introduced to quantify the expressivity. While Expr provides insights into a QNN’s ability to explore the Hilbert space, its computation requires extensive sampling over the entire parameter space, similar measures such as the effective dimension~\cite{abbas_power_2021} and effective capacity~\cite{wilkinson_evaluating_2022, maronese_high-expressibility_2026}, making it computationally expensive. 
Additionally, Expr is defined as the Kullback–Leibler divergence between the estimated fidelity distribution and that of the Haar-distributed ensemble, restricting its applicability to unitary circuits and making it unsuitable for non-unitary architectures such as dissipative QNNs~\cite{beer_training_2020}. 
Other studies have assessed expressivity using Fourier analysis~\cite{schuld_effect_2021}, entangling power~\cite{hubregtsen_evaluation_2021, ballarin_entanglement_2023}, 
{covering number~\cite{du_efficient_2022}} 
or learning performance~\cite{wu_expressivity_2021,wu_randomness-enhanced_2024}. {However, these measures are typically defined at the level of unitary operations or specific encoding strategies and do not directly account for the interplay between the input dataset, circuit structure and measurement protocol.}

{In parallel, Fisher-information-based approaches have provided an alternative perspective by relating the geometry and rank of the Fisher information matrix to circuit capacity and trainability~\cite{haug_capacity_2021,meyer_fisher_2021}. However, these studies mainly treat Fisher information as a diagnostic tool and often focus on specific aspects of QNN design in isolation. A systematic framework that captures the joint dependence of expressivity on data encoding, measurement protocol, and circuit architecture remains lacking.
Despite these advances, the absence of a unified and generalizable framework for quantifying and optimizing QNN expressivity remains a central challenge. To address this, we develop a Fisher-information-based framework in which the effective rank ($\kappa$) serves as a central quantity for characterizing circuit expressivity~\cite{liu_quantum_2019, meyer_fisher_2021, Karakida_2020,haug_capacity_2021}. }

{In contrast to prior works that primarily rely on the quantum Fisher information~\cite{meyer_fisher_2021, RevModPhys.90.035005}, we focus on the classical Fisher information matrix associated with specific measurement protocols. This choice allows us to explicitly incorporate the role of measurement into the analysis, which is essential for practical quantum learning tasks and enables a direct connection between expressivity and experimentally accessible quantities. 
Within this framework, $\kappa$ captures the number of effectively independent parameters in a quantum circuit, providing a computationally tractable and physically interpretable characterization of its expressive capacity. Importantly, this formulation naturally reveals the joint dependence of expressivity on data encoding, measurement protocol, and circuit architecture, thereby establishing a unified perspective. Furthermore, we show that $\kappa$ can be promoted from a diagnostic quantity to a practical design objective, enabling the optimization of circuit architectures in reinforcement-learning-based approaches.}

\section{Methods}
\begin{figure}[t]
\begin{centering}
\includegraphics[width=0.85\textwidth]{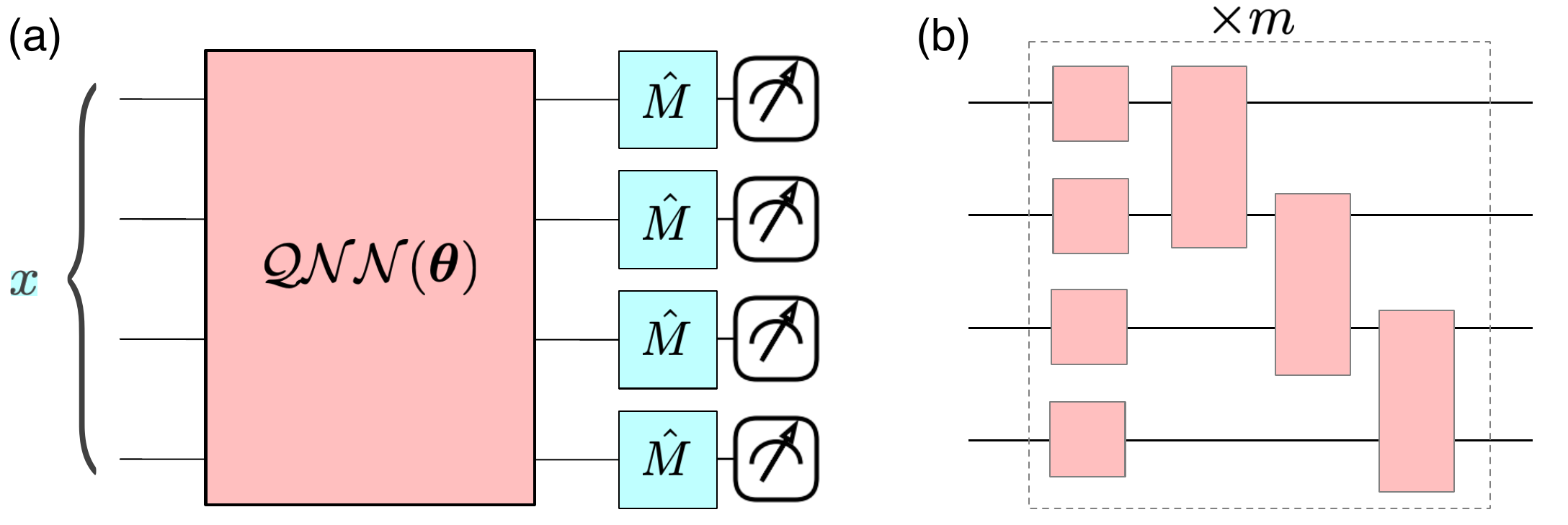}
\caption{Schematic of a quantum neural network. 
(a) A quantum neural network with an input quantum state $x$ and the measurement basis operator $\hat{M}$ for a four-qubit system ($n=4$). (b) The building block comprises  $n$ single-qubit gates and $n-1$ two-qubit gates, arranged in a chain configuration.  Multiple building blocks are used to construct the quantum neural network depicted in (a).  }
\label{Fig1}
\end{centering}
\end{figure}

{As shown in Fig.~\ref{Fig1}(a), a quantum neural network (QNN) is constructed from a parameterized quantum circuit (PQC) that processes an input quantum state. Given an input $x$, a corresponding quantum state $\rho(x)$ is prepared via a data-encoding procedure, followed by a parameterized unitary transformation $U({\bm \theta})$ defined by the circuit ansatz, resulting in
\begin{equation}
\rho_{\bm \theta}(x) = U({\bm \theta}) \rho(x) U^\dagger({\bm \theta}).
\end{equation}
Measurements described by a set of operators $\{\hat{M}\}$ are then performed on the evolved state, giving rise to the conditional probability distribution
\begin{equation}
P(y|{\bm \theta}, x) = \mathrm{Tr}\left[ \hat{M} \rho_{\bm \theta}(x) \right],
\end{equation} 
Here $y$ denotes the measurement outcome bit string, which occurs with probability $P(y|{\bm \theta}, x)$. }

{Treating the QNN as a statistical model, the Fisher information matrix (FIM) is defined as
\begin{equation}
F_{ij}({\bm \theta}) \equiv \mathbb{E}_{(x, y)\sim P}\left[
\frac{\partial}{\partial \theta_i}\log P(y|{\bm \theta},x)
\frac{\partial}{\partial \theta_j}\log P(y|{\bm \theta},x)
\right],
\end{equation}
which quantifies the sensitivity of the output distribution with respect to the parameters ${\bm \theta}=\{\theta_j\}_{j=1}^p$. The expectation is taken over the input distribution and measurement outcomes. 
From this formulation, it is clear that the structure of the Fisher information matrix is jointly determined by three key components: the data encoding $\rho(x)$, the circuit ansatz $U({\bm \theta})$, and the measurement operators $\{\hat{M}\}$. This dependence forms the basis for our unified analysis of QNN expressivity.}

{In the following, we use the rank of the Fisher information matrix as a metric for QNN expressivity, which we refer to as the effective rank, $\kappa$.
The Fisher information matrix is a $p \times p$ positive semidefinite matrix with non-negative eigenvalues. In practice, its rank can be reliably estimated using randomly sampled parameter values, except for a set of measure-zero singular configurations.
Compared to other metrics such as Expr or the effective dimension, the effective rank does not require extensive sampling over the parameter space and is therefore computationally more efficient.
We further show that $\kappa$ captures the number of effectively independent directions in the parameter space, providing an estimate of the intrinsic degrees of freedom of the quantum circuit.
Within this framework, we systematically investigate how data encoding $\rho(x)$, circuit ansatz $U({\bm \theta})$, and measurement operators $\{\hat{M}\}$ jointly influence the expressivity of QNNs through the effective rank, thereby establishing a unified characterization.}

\section{Results}
\subsection{Influence of Data and Measurement on QNN Expressivity  }
As illustrated in Fig.~\ref{Fig1} (a), the Fisher information matrix of a quantum neural network (QNN) is shaped  by three key factors: the quantum circuit architecture, input quantum states, and measurement protocol.
Ref.~\cite{haug_capacity_2021} proposed that the rank of the quantum Fisher information matrix can serve as an indicator of a quantum circuit capacity. However, this analysis was restricted to a fixed pure input state and neglected the influence of the measurement protocol.
In practice, a quantum neural network acts as a variational model that represents input quantum states, whereas the choice of input states or training dataset simultaneously imposes constraints on the network’s trainable parameters.
Consequently, the configuration of both the input states and the measurement protocol plays a decisive role in determining the expressivity of the QNN. To verify this hypothesis, we conducted systematic numerical analyses based on an effective rank metric. By varying the configuration of the input quantum states and measurement operators, we reveal that the expressivity of the quantum neural network is not an intrinsic property of the circuit alone, but rather emerges from the interplay between data encoding and the measurement protocol.

To eliminate the influence of circuit architecture on expressivity, we consider a universal parameterized quantum circuit defined as
\begin{equation}
\mathcal{U}({\bm \theta})=e^{i \sum_{j=1}^{d_n} \theta_j \hat{g}_j},
\label{EqUuni}
\end{equation}
where $\hat{g}_j$ are the generators of the $SU(2^n)$ group, and $\theta_j$ are the variational parameters. Index $g$ runs over all independent generators, with $d_n = 4^n-1$ which corresponds to the dimension of the Lie algebra associated $SU(2^n)$.  In this case, the variational parameters were formally independent. However, owing to the restricted set of input quantum states and measurement observables, only a subset of these parameters plays an effective role. Consequently, the number of effective independent parameters, as captured by the effective rank, is generally smaller than $d_n$. Only when the input data and measurement protocol are sufficiently informative can the effective rank reach its maximal value of $d_n$, as discussed in detail below. 

\begin{figure}[t]
\begin{centering}
\includegraphics[width=0.75\textwidth]{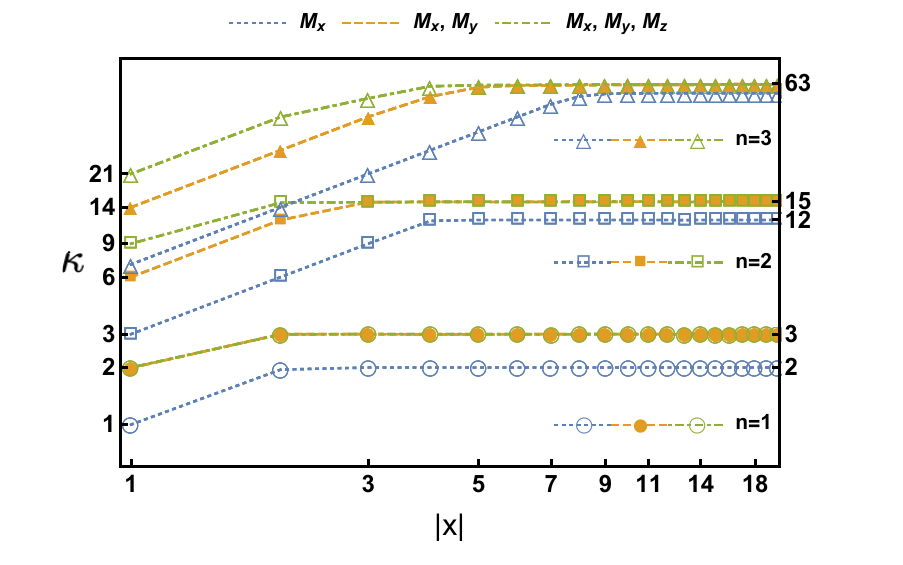}
\caption{The effective ranks, $\kappa$, are plotted as a function of the number of input quantum states, $|x|$, for qubit numbers $n=1, 2, 3$, represented by circular, square, and triangular markers, respectively.
Results for measurement protocols with one $(M_x)$, two $(M_x, M_y)$, and three $(M_x, M_y, M_z)$ basis operators are shown as dotted, dashed, and dot-dashed lines, respectively. }
\label{Fig2}
\end{centering}
\end{figure}

As shown in Fig.~\ref{Fig2}, the effective rank $\kappa$ is presented as a function of the total number of input quantum states for different measurement protocols.
The measurement protocol is defined by the set of basis operators applied prior to the projection measurement, as illustrated in Fig.~\ref{Fig1}(a). $M_j$ denotes the basis operator corresponding to the projection along the $j$-direction.
In our calculations, each input quantum state is randomly generated as a valid density matrix 
{
using the Wishart construction \cite{PhysRevA.102.012405, codelink}. 
The numerical simulations  are based on the Pennylane library interfaced with PyTorch~\cite{bergholm_pennylane_2022}.
}
For single-projection measurements along $x$-basis, as shown by the blue dashed lines in Fig.~\ref{Fig2}, 
$\kappa$ starts at a low value and increases monotonically toward its upper bound for all qubit numbers $n=1,2,3$,
{where the full Fisher information matrix can be explicitly computed. This saturation behavior becomes computationally inaccessible for larger $n$, as it would require variational circuits approaching maximal expressibility.}

When a sufficient number of input quantum states are reached, adding more data samples to the dataset no longer increases the effective rank, although the effective rank is still lower than $d_n$. 
{Here, the effective rank $\kappa$ quantifies the number of linearly independent parameter directions that influence the model output, and thus characterizes the number of independent (non-redundant) variational parameters in the circuit.}
This behavior indicates that increasing the number of input quantum states imposes stronger constraints on the variational parameters, reduces redundancy, and enhances their effective independence. However, this enhancement only occurs if the complexity of the dataset is not saturated. Beyond a critical point, adding more data does not further increase $\kappa$, indicating that no additional independent parameter directions can be resolved, and thus the expressivity of the QNN is not further improved.

Nevertheless, incorporating additional observables can extend the upper bound imposed by the input dataset. When more observables are included, as indicated by the orange and green lines in Fig.~\ref{Fig2}, the effective rank can reach the maximum value $4^n-1$, which corresponds to the number of generators in the Lie algebra of the $n$-qubit system. At this point, all the $4^n-1$ variational parameters $\theta_j$ correspond to independent directions, and the quantum neural network achieves its maximal expressivity.

The enhancement of the effective rank through additional input states and observables can be intuitively understood by analogy with solving a system of equations: increasing the number of input samples or observables is akin to adding more independent equations, thereby providing additional information for parameter estimation, resolving previously redundant parameter directions, and reducing degeneracies in the parameter space. {This interpretation is consistent with the rank of the Fisher information matrix, which captures the number of locally identifiable parameter directions.}

In this section, we demonstrate that both the input dataset and measurement protocols can affect the expressivity of the QNNs. Formally independent variational parameters can effectively implement a function only when the input data and measurement protocols are sufficiently informative.

\subsection{Influence of Circuit Architecture on QNN Expressivity}
The universal parameterization of the quantum circuit, as described in Eq.~\eqref{EqUuni}, is generally impractical because the number of formally independent parameters $d_n$ grows exponentially with the number of qubits. 
In realistic implementations, hardware-efficient quantum circuits are constrained by limited qubit connectivity and shallow depth, preventing them from fully exploring the Hilbert space in Eq.~\eqref{EqUuni}. Thus, the effective rank serves as a practical measure of expressivity by quantifying the number of effective independent parameters realized by a given circuit architecture. In general, hardware-efficient designs exhibit effective ranks that are far below the theoretical maximum, particularly for large-scale systems.

\begin{figure}[t]
\begin{centering}
\includegraphics[width=0.75\textwidth]{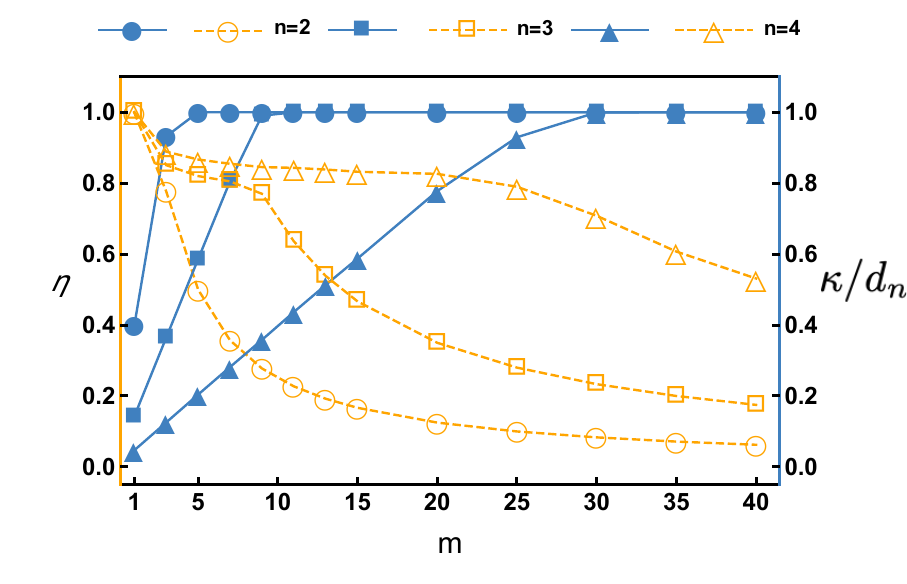}
\caption{For $n=2, 3, 4$,  the effective ranks of the quantum neural network relative to the upper bound, $\kappa/d_n$, are plotted by the blue solid lines, while the parameter efficiency, $\eta = \kappa/p$, is shown by the orange dashed lines. 
Here, $m$ represents the number of building blocks in the quantum circuit, and $d_n=4^n-1$ denotes the optimal number of independent parameters for an $n$-qubit unitary transformation. }
\label{Fig3}
\end{centering}
\end{figure}
For demonstration purposes, we investigated how the architecture of a quantum circuit affects its expressivity, using the effective rank as the characterizing metric. In the hardware-efficient configuration illustrated in Fig.~\ref{Fig1}(b), a circuit is constructed by stacking $m$ copies of a basic building block. Each building block follows a chain pattern and includes nearest-neighbor two-qubit CNOT gates, as well as single-qubit rotations parameterized by three variational parameters, $\phi, \omega,$ and $\theta$:
\begin{equation}
\hat{u}(\phi,\omega,\theta) =
\begin{bmatrix}
\cos\frac{\theta}{2}e^{-i\frac{\phi+\omega}{2}} & -\sin\frac{\theta}{2}e^{i\frac{\phi-\omega}{2}} \\
\sin\frac{\theta}{2}e^{-i\frac{\phi-\omega}{2}} & \cos\frac{\theta}{2}e^{i\frac{\phi+\omega}{2}}
\label{Eq5}
\end{bmatrix}.
\end{equation}
To isolate the effect of the circuit architecture, a sufficiently rich set of input states and measurement operators is employed so that the effective rank is not constrained by the dataset or measurement protocol. The resulting effective rank $\kappa$ is then evaluated as a function of the circuit depth $m$, as indicated by the blue lines in Fig.~\ref{Fig3}.

With the addition of more building blocks, the effective rank $\kappa$ increased monotonically, reflecting the enhancement of expressivity achieved by stacking additional layers. However, increasing the circuit depth introduces more variational parameters. The ratio of $\kappa$ to the total number of parameters $p$, denoted as $\eta$, quantifies the parameter efficiency of the circuit. As shown by the orange lines in Fig.~\ref{Fig3}, adding more building blocks increases expressivity but simultaneously reduces parameter efficiency. A smaller $\eta$ corresponds to more eigenvalues of the Fisher information matrix approaching zero, which can result in the barren plateau phenomenon~\cite{abbas_power_2021}. 
Therefore, the parameter efficiency $\eta$ is closely linked to the trainability of the circuit~\cite{PRXQuantum.2.040337, qi_theoretical_2023, PRXQuantum.3.030323}.
{For a fixed effective rank $\kappa$, increasing $\eta$ improves the trainability of the model. Conversely, a smaller $\eta$, which indicates higher redundancy, gives rise to more eigenvalues of the Fisher information matrix approaching zero. As a result, the gradients along these directions become negligible, reducing the amount of useful information available for parameter updates and consequently making the optimization more challenging.}

These observations revealed  an inherent trade-off between circuit expressivity and parameter efficiency. Specifically, increasing the circuit depth and the number of variational parameters enhances the expressivity but often reduces the parameter efficiency. This trade-off highlights the need for a systematic design strategy that can balance expressivity and parameter economy through an efficient allocation of variational parameters. Guided by the effective rank, in the following, we employed a reinforcement learning algorithm to automatically construct quantum circuits with finite depth and a limited number of variational parameters.

\subsection{Architecture Design via Reinforcement Learning}
{The above results show that the expressivity of a QNN is governed by the interplay among the dataset, measurement operators, and circuit structure. 
Here, we focus on circuit design with a fixed dataset and measurement scheme. 
As a quantitative measure of expressivity, the effective rank will be employed to guide the circuit design via a reinforcement learning framework.
}

\begin{figure}[t]
\begin{centering}
\includegraphics[width=0.75\textwidth]{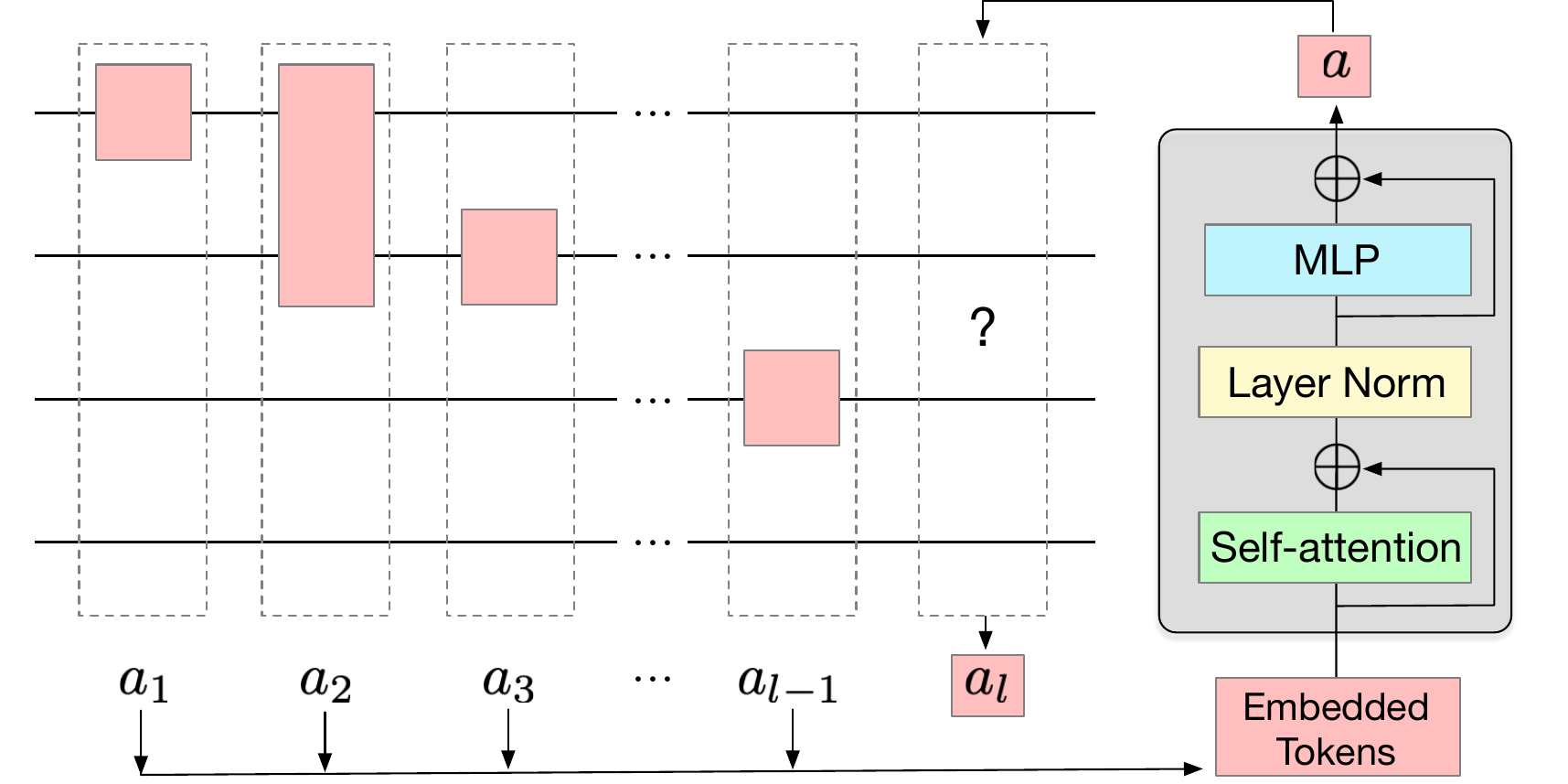}
\caption{ 
{Sequential quantum circuit design process. 
The gate at layer $l$, denoted by $a_l$,  is produced by a self-attention transformer agent conditioned on the previous gate sequence $\{a_1, a_2, \dots, a_{l-1}\}$. The complete circuit of depth $L$ is obtained by iteratively querying the agent. For each iteration, the reinforcement learning state is denoted as $\{a_1, a_2, \dots, a_{l-1}\}$, the action is $a_l$. The corresponding reward $R(a_l^s | a_{1:l-1}^s)$ is the effective rank of the partial circuit denoted as $\{a_1, a_2, \dots, a_{l}\}$. 
}
}
\label{Fig4}
\end{centering}
\end{figure}

{As shown in Fig.~\ref{Fig4}, the circuit design can be viewed as a generative process. A quantum circuit is incrementally constructed gate-by-gate. In each step, the available quantum gates include parameterized single-qubit gates (Eq.\eqref{Eq5}) and two-qubit CNOT gates with arbitrary qubit connections.
For an $n$-qubit quantum circuit, the number of available gate operations per step is n + n(n-1), 
where the first term corresponds to single-qubit gates and the second term corresponds to two-qubit CNOT gates acting on ordered control–target qubit pairs.
The total number of gate operations is $n^2$. We tokenize the gate operations by encoding both the gate type and the target qubit index, as shown in Tab.~\ref{tableID}. 
To unify the representation, we index gate operations by a pair $(i,j)$. 
Single-qubit gates correspond to the diagonal terms $(i=j)$, while two-qubit CNOT gates correspond to off-diagonal terms $(i\neq j)$. Each pair $(i,j)$ defines a unique token, resulting in a total of $n^2$ tokens, indexed by a token ID $a \in \{0,\dots,n^2-1\}$.
}
\begin{table}[h]
\centering
\begin{tabular}{c|c|c}
\hline
 & Single-qubit & Two-qubit (CNOT) \\
\hline
Qubit index 
& $(i,j),\, i=j$ 
& $(i,j),\, i\neq j$ \\
\hline
Token ID ($a$)
& \multicolumn{2}{c}{~~$n i + j$ ~~($i,j \in \{0,\dots,n-1\}$)} \\
\hline
\end{tabular}
\caption{ {Tokenization of gate operations using a unified index pair $(i,j)$. }}
\label{tableID}
\end{table}

{
The construction process begins with a reinforcement learning (RL) state representing an initial or partially constructed circuit, which is encoded as a sequence of gate tokens $a_1,a_2,a_3,...a_{l-1}$. This state serves as the input to the agent, whose output determines the next action $a_l$, as schematically illustrated in Fig.~\ref{Fig4}. Each query to the agent extends the sequence and thereby updates the circuit configuration. This procedure is iteratively repeated until the circuit reaches a predefined length $L$. Note that the circuit length $L$, defined as the total number of sequentially applied gate operations, differs from the conventional notion of circuit depth, which is given by the length of the longest path of gate layers that can be executed in parallel.
Without loss of generality, we set the initial action as $a_1=0$, representing a  single-qubit gate acting on the first qubit.
}

{
As shown in the right panel of Fig.~\ref{Fig4}, the agent is parameterized as a self-attention transformer~\cite{vaswani_attention_2017}. 
For a circuit of depth $L$, the number of possible circuit configurations scales as $n^{2L}$, leading to an exponentially large search space. 
This exponential scaling renders exhaustive search computationally intractable. 
To overcome this challenge, we train the agent within a reinforcement learning framework to efficiently generate quantum circuit configurations using a policy gradient method.
This efficiency arises because the agent learns a policy that sequentially generates high-quality gate sequences based on the preceding configuration, guided by a reward signal. Instead of exhaustively enumerating all 
$n^{2L}$ possible circuits, the reinforcement learning framework focuses the search on promising configurations, thereby reducing computational cost while effectively exploring the exponentially large circuit space.
In our framework, the effective rank of the quantum circuit serves as the reward for the reinforcement learning agent. 
}

{
The reinforcement learning model is initialized with randomly sampled transformer parameters. 
In the first iteration, $10$ circuit configurations, each represented as a sequence $\{a_j\}_{j=1}^L$, are randomly generated to form the initial training dataset. 
The transformer is optimized using a policy gradient loss function defined as
\begin{equation}
\mathcal{L} = -\frac{1}{\mathcal{N}L} \sum_s \sum_{l=2}^L \log P(a_l^s | a_{1:l-1}^s)\, R(a_l^s | a_{1:l-1}^s),
\end{equation}
where $s$ indexes the sampled trajectories and $\mathcal{N}$ is the total number of samples. 
Here, $P(a_l^s | a_{1:l-1}^s)$ denotes the probability assigned by the transformer to action $a_l^s$, 
and $R(a_l^s | a_{1:l-1}^s)$ is the corresponding reward, defined as the effective rank $\kappa$ of the partial circuit $\{a_1, \dots, a_l\}$. 
After each training iteration, $10$ new circuit sequences are generated by the updated policy to augment the dataset. 
This sampling and training procedure is repeated until a circuit configuration achieving a sufficiently large effective rank $\kappa_{\max}$ is found.
}

\begin{figure}[t]
\begin{centering}
\includegraphics[width=0.75\textwidth]{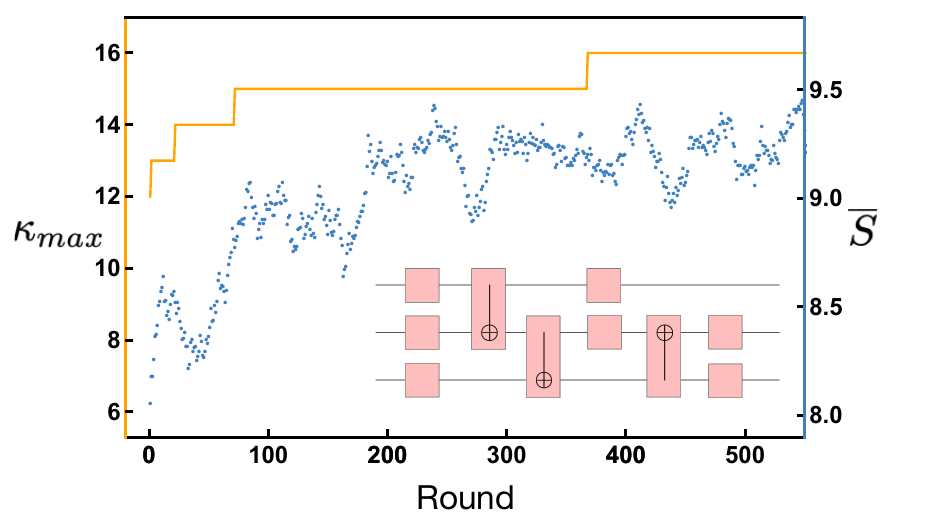}
\caption{ 
Reinforcement learning process for constructing a quantum circuit with $n=3$ qubits and depth $L=10$. In each round, $10$ configurations are generated. The orange curve represents the maximum effective rank $\kappa_{max}$ among all configurations generated up to the round indicated on the x-axis. The blue dots indicate the average effective rank over a sliding window of $10$ rounds, defined as the score $\overline{S}$ of the agent.
   }
\label{Fig5}
\end{centering}
\end{figure}
{For demonstration, we consider an $N=3$ qubit system with a circuit depth of $L=10$, given a fixed quantum state dataset and measurement protocol. 
The reinforcement learning model is implemented in PyTorch and trained using the Adam optimizer~\cite{codelink}. }
The performance of the trained agent and the resulting neural network architecture are shown in Fig.~\ref{Fig5}. 
{In each round, $10$ configurations are generated. The orange curve represents the maximum effective rank $\kappa_{max}$ among all configurations generated up to the round indicated on the x-axis. The blue dots indicate the average effective rank over a sliding window of $10$ rounds, defined as the score $\overline{S}$ of the agent.}
The average effective rank $\bar{S}$, represented by the blue dots, exhibits short-term fluctuations owing to the stochastic nature of the training process, such as the exploration by the agent. Nonetheless, the overall upward trend of $\bar{S}$ indicates a steady improvement in the agent’s ability to generate expressive circuits.

As shown by the orange line,  {the maximum effective rank $\kappa_{\max}$ reaches 16 within around 400 training rounds, significantly exceeding both the value of 13 for a chain configuration of the same depth and that obtained by random search over thousands of rounds}.
The identified optimal configuration is shown in the inset of Fig.~\ref{Fig5}. Compared with the chain structure, a two-qubit gate in the second block is replaced by an additional single-qubit gate. Despite having fewer two-qubit gates, the circuit exhibits enhanced expressivity, as evidenced by its larger effective rank. This reduction is also advantageous for experimental implementation, because single-qubit gates typically offer higher fidelity and lower error rates. Overall, the reinforcement-learning-based design strategy yields more expressive and experimentally feasible architectures than the conventional manually designed ones.

\section{Discussion and Conclusion}
In general, a quantum neural network is constructed by applying a series of variational parameters embedded in different types of quantum gates. The independence of these parameters is essential for the model’s expressivity because redundant parameters do not contribute to an increase in the network’s expressive power.
We found that the independence of the variational parameters can be effectively quantified by the rank of the Fisher information matrix.

We introduced the effective rank $\kappa$, as a quantitative measure of QNN expressivity.
The effective rank offers a computationally tractable yet physically meaningful characterization of expressivity.
Using $\kappa$, we systematically investigated the influence of the input dataset, measurement protocol, and circuit architecture on expressivity.
We demonstrate that $\kappa$ can reach its theoretical upper bound, $d_n$, for an $n$-qubit system through appropriate design of these components.
Thus, $\kappa$ serves as a direct estimate of the number of  independent parameters within a quantum circuit, offering a rigorous and scalable framework for quantifying expressive capacity.

Building upon $\kappa$, we developed a reinforcement learning framework based on a self-attention transformer agent to autonomously optimize quantum circuit configurations.
{The scalability to larger qubit systems and longer circuit depths requires more advanced algorithmic developments, such as more efficient reward evaluation strategies, improved credit assignment mechanisms, and reinforcement learning architectures tailored to structured sequence generation problems.
}
{The proposed framework is directly compatible with near-term quantum hardware, as it relies on parameterized quantum circuits and standard measurement protocols. For a given experimental setting with fixed connectivity, measurement schemes, and circuit depth, our approach enables the systematic design of optimized circuit configurations tailored to these constraints.
The optimization can be performed offline via classical simulations, allowing efficient exploration of the circuit space without requiring access to quantum hardware. This significantly reduces experimental overhead and facilitates practical implementation. 
Moreover, to mitigate the impact of noise and measurement imperfections in real devices, the optimization can also be extended to an online setting. In this scenario, the reinforcement learning procedure can be implemented within a hybrid quantum-classical loop, where circuit evaluations are performed on quantum hardware while the policy is updated classically.
These considerations demonstrate the feasibility of our approach in realistic settings and highlight its potential for applications on current and near-term quantum platforms.}

{
By covering all three stages of a QNN (input data, parameterized quantum circuit, and measurement), the effective rank provides a unified and comprehensive characterization of QNN expressivity. 
Beyond constructing high-performance circuit configurations, the effective rank may also provide a useful framework for evaluating different strategies for embedding input data into the Hilbert space of qubits, as well as measurement protocols for extracting information from selected observables. 
Exploring these directions constitutes an interesting avenue for future work.}

\section{Acknowledgements}
We thank Augusto Smerzi, Georg Engelhardt, Yuqi Zhang and  Hui Zhai for helpful discussions.
This work is supported by the Science, Technology and Innovation Commission of Shenzhen, Municipality (KQTD20210811090049034), Guangdong Basic and Applied Basic Research Foundation (2022B1515120021), and  
National Natural Science Foundation of China (Grant No. 11904190).

\section{Data availability statement}
The datasets are available at GitHub repository for Learning to Maximize Quantum Neural Network Expressivity via Effective Rank \cite{codelink}.

\section{Conflicts of interest}
The authors declare that they have no competing financial interests.

\bibliographystyle{unsrtnat}

\end{document}